\documentclass[nofootinbib,twocolumn,a4paper,floatfix,prl]{revtex4}
\usepackage{graphicx}
\usepackage{amsfonts}
\usepackage{amssymb}
\usepackage{amsmath}
\usepackage{latexsym}
\usepackage{subfloat}

\begin{document}

\preprint{ICG preprint 05/xx}

\title{Delocalization of brane gravity by a bulk black hole}

\author{Sanjeev S.~Seahra}
\author{Chris Clarkson}
\author{Roy Maartens}
\affiliation{Institute of Cosmology \& Gravitation, University of
Portsmouth, Portsmouth, PO1 2EG, UK}

\setlength\arraycolsep{1pt}
\newcommand*{\di}{\partial}
\newcommand*{\ds}[1]{ds^2_\text{\tiny{($#1$)}}}
\newcommand*{\ads}[1]{{AdS$_{#1}$}}
\newcommand*{\rb}{r_\text{b}}
\newcommand*{\xb}{x_\text{b}}
\newcommand*{\Rh}{R_\text{h}}
\newcommand*{\fb}{f_\text{b}}

\date{\today}

\begin{abstract}

We investigate the analogue of the Randall-Sundrum brane-world in
the case when the bulk contains a black hole. Instead of the
static vacuum Minkowski brane of the RS model, we have an Einstein
static vacuum brane. We find that the presence of the bulk black
hole has a dramatic effect on the gravity that is felt by brane
observers. In the RS model, the 5D graviton has a stable localized
zero-mode that reproduces 4D gravity on the brane at low energies.
With a bulk black hole, there is no such solution
--- gravity is delocalized by the 5D horizon. However, the brane
does support a discrete spectrum of metastable massive bound
states, or quasinormal modes, as was recently shown to be the case
in the RS scenario. These states should dominate the high
frequency component of the bulk gravity wave spectrum on a
cosmological brane.  We expect our results to generalize to any
bulk spacetime containing a Killing horizon.

\end{abstract}

\maketitle

\subsection*{Introduction}

The Randall-Sundrum (RS) model~\cite{rs} consists of a 4D
Minkowski brane embedded (with mirror symmetry) in a 5D anti de
Sitter (AdS) bulk spacetime. Although the extra dimension is
infinite, it is exponentially warped and this leads to the
recovery of 4D gravity on the brane at low energies. Thus the RS
model provides the basis for models that could describe the
observed universe as a brane-world. The key to this feature is the
recovery of General Relativity at low energies, i.e., the
existence of a normalizable zero-mass mode of the 5D graviton that
is a bound state on the brane. General Relativity acquires
corrections from the massive modes of the 5D graviton on the
brane. (The massive modes dominate over the zero-mode at high
energies.)


Why does the RS model allow for a brane-localized zero-mode? The
reason is the warp-factor, which erects a potential barrier around
the brane that efficiently ``squeezes'' bulk gravity waves. In
this work, we explore the consequences of modifying this barrier
by introducing Weyl curvature into the model via a bulk black
hole. In order to have a static vacuum configuration, we require a
curved brane with an Einstein-static metric.  The black hole's
presence has a drastic implication for the zero-mode: there is no
longer a gravity-wave bound state solution, because the extreme
gravitational field near the horizon causes the brane potential
barrier to become highly permeable, or ``leaky''.

The bulk metric satisfies the 5D Einstein equations
$G^{(5)}_{ab}=-\Lambda_5 g^{(5)}_{ab}$, and is given by
\begin{eqnarray}
    ds^2_{(5)} & = & -f\,dT^2 + f^{-1}
    \,dR^2 + R^2 \, d\Omega_{(3)}^2\,,\label{met} \\
    f & = & 1 -
    \frac{R_0^2}{R^2} +
    \frac{R^2}{\ell^2}= \frac{(R^2 + \Rh^2+\ell^2)(R^2 -
    \Rh^2)}{R^2 \ell^2},\label{sads}
\end{eqnarray}
where $R_0$ determines the ADM mass of the black hole, $\ell$ is
the AdS length scale defined by the bulk cosmological constant
($\Lambda_5=-6/\ell^2$), and
\begin{equation}
    \Rh^2 = \frac{\ell^2}{2} \left(- 1+ \sqrt{
    \frac{4R_0^2}{\ell^2} + 1}  \right).
\end{equation}
Thus, there is an event horizon at $R = \Rh$.  We define
dimensionless coordinates $(t,r)=(T/\Rh,R/\Rh)$, so that
\begin{eqnarray}
    f(r) = \frac{(r^2 + \gamma^2 + 1)(r^2-1)}{\gamma^2 r^2}, \quad
    \gamma \equiv \frac{\ell}{\Rh}.
\end{eqnarray}
The bulk geometry is completely characterized by the ratio of the
AdS length scale to the black hole horizon radius, $\gamma$, and
the horizon is always at $r = 1$. It is sensible to call solutions
with $\gamma \ll 1$ ``big'' black holes and solutions with $\gamma
\gg 1$ ``small'' black holes.

A static brane is introduced by identifying $r=\rb$ as the
boundary of the bulk and discarding the region $r > \rb$. The
brane metric has the Einstein static form,
\begin{equation}\label{es}
    ds^2_{(4)} = -d\tau^2 + \Rh^2 \rb^2 \, d\Omega_{(3)}^2\,,
\end{equation}
where $\tau = \fb \Rh t = \fb T$ is the cosmic time. In the RS
model, the tension (vacuum energy) of the brane exactly cancels
the bulk cosmological constant so that the effective 4D
cosmological constant vanishes and the brane is Minkowski. Here,
the tension exceeds the critical RS value, so that the brane has
positive cosmological constant.  The acceleration due to this
effective cosmological constant is nullified by the ``dark
radiation'' induced by the projection of the bulk Weyl curvature
onto the brane. Hence, unlike General Relativity we do not need
any additional matter to have an Einstein-static configuration
\cite{gm}. This is a delicate balance, which is why the brane's
position must be fine-tuned.

We can verify this via the junction conditions. The jump in the
extrinsic curvature is determined by the energy and stresses on
the brane. By the standard junction conditions and the mirror
symmetry, the extrinsic curvature of a brane with tension $\sigma$
must satisfy:
\begin{equation}
K_{ab} \equiv g^c{}_b \nabla^{(5)}_c n_a = -\frac{\kappa_5^2}{6}\,
\sigma g_{ab}\,, \quad g_{ab} \equiv g^{(5)}_{ab} - n_a n_b\,,
\end{equation}
where $n_a = -f^{-1/2} \delta_a{}^R$ is the normal. Together with
Eq.~(\ref{es}), this leads to two equations in the three
parameters $\gamma$, $\sigma$, and $r_b$, with solutions
\begin{equation}\label{esp}
\sigma(\gamma) = \frac{\gamma^2+2}{2\gamma\sqrt{\gamma^2+1}},
\quad \rb(\gamma) = \frac{\sqrt{2\gamma^2+2}}{\gamma}.
\end{equation}
The solution for $\rb$ shows that a pure tension brane is
coincident with the photon-sphere of the bulk black hole. (This is
a special case of the general result that pure tension branes are
always totally geodesic for null paths, i.e., they are umbilical
surfaces.) Equations~(\ref{met}), (\ref{es}) and (\ref{esp})
define the Einstein-static (ES) brane-world model.

\subsection*{Tensor perturbations}

We now consider tensor perturbations of the bulk metric
\begin{equation}
\ds{5} = \Rh^{2}[-f\,dt^2 + f^{-1}\,dr^2 + r^2 ( h_{ij} + \delta
h_{ij} ) d\theta^i d\theta^j]\,,
\end{equation}
where $h_{ij}d\theta^i d\theta^j=d\Omega_{(3)}^2$. We harmonically
decompose the metric perturbation,
\begin{equation}
\delta h_{ij} = \sum_k {r^{-3/2}}{\psi_k(t,r)}
\mathbb{T}^{(k)}_{ij}\,.
\end{equation}
The tensor harmonics are defined by
\begin{equation}
\roarrow{\nabla}^2 \mathbb{T}^{(k)}_{ij} = -k^2
\mathbb{T}^{(k)}_{ij}\,, \quad \roarrow{\nabla}^i
\mathbb{T}^{(k)}_{ij} = 0 = h^{ij} \mathbb{T}^{(k)}_{ij}\,,
\end{equation}
where $\roarrow{\nabla}_i$ is the covariant derivative of $h_{ij}$
and
\begin{equation}
    k^2 = L(L+2)-2\,, \quad L = 1,2,3,\ldots
\end{equation}
Using the general results of Kodama and
Ishibashi~\cite{Kodama:2003jz} specialized to 5D
Schwarzschild-AdS, we see that the linearized Einstein equations
show that $\psi_k$ satisfies a wave equation,
\begin{eqnarray}
-\frac{\di^2 \psi_k}{\di t^2} & = & - \frac{\di^2 \psi_k}{\di
x^2} + V_k(r)\psi_k, \quad x \equiv \int \frac{dr}{f}\,,
\label{master wave pde} \\
V_k(r) & = & f \left[ \frac{15}{4\gamma^2} + \frac{4 k^2 + 11}{4
r^2} + \frac{9 (\gamma^2+1)}{4\gamma^2 r^4} \right]\!,
\end{eqnarray}
where $x$ is the tortoise coordinate and the bulk black hole
horizon is at $x=-\infty$. The boundary condition follows from the
requirement that the bulk fluctuation preserves the matter content
of the brane, i.e., the brane has only tension after perturbation:
$\delta K_{ab} = -\kappa_5^2\sigma \, \delta g_{ab}/6$, which
implies
\begin{equation}\label{boundary condition}
\left[\di_r( r^{-3/2} \psi_k) \right]_{\rm b}=0\,.
\end{equation}

It is useful to compare Eq.~(\ref{master wave pde}) with the
analogous master wave equation in the one-brane RS model:
\begin{eqnarray}
-\frac{\di^2 \phi_k}{\di t^2} & = & - \frac{\di^2 \phi_k}{\di z^2}
+ U_k(z)\phi_k\,,\label{RS master wave pde} \\\ U_k(z) & = & k^2 +
\frac{15}{4(z-2)^2}\,,
\end{eqnarray}
where $z$ is a dimensionless conformal coordinate. (For ease of
comparison, we are using the mirror image of the bulk on the left
of the brane, $z_{\rm b}=1$.) Since the spatial geometry is flat
in the RS scenario, $k$ is a continuous parameter. The RS boundary
condition is
\begin{equation}
\left[ \di_z  (z^{-3/2} \phi_k )\right]_{\rm b}=0\,.
\end{equation}
\begin{figure}
\begin{center}
    \includegraphics{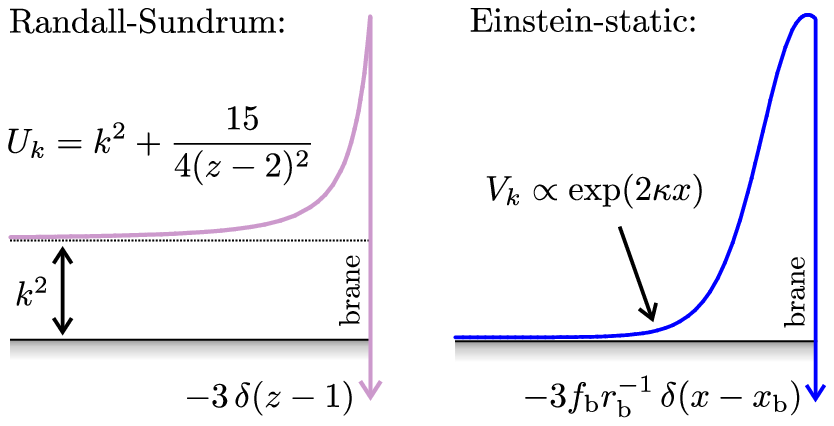}
\end{center}
\caption{Comparison of the tensor potentials in the
Randall-Sundrum and Einstein-static
brane-worlds.}\label{fig:comparison}
\end{figure}

The $U_k$ and $V_k$ potentials are sketched in
Fig.~\ref{fig:comparison}, where we have added a delta function at
the brane position to enforce the boundary condition. The major
difference between the two potentials is their asymptotic
behaviour far from the branes. In the RS case, the potential
decays like $1/z^2$ to a positive constant. By contrast, the ES
potential decays as ${\rm e}^{2\kappa x}$ for $x \rightarrow
-\infty$ (the black hole horizon), where $\kappa$ is the
dimensionless surface gravity:
\begin{equation}
    \kappa = \frac{1}{2} f'(1) =
    \frac{2+\gamma^2}{\gamma^2}\,.
\end{equation}

\subsection*{Delocalization of the zero-mode}

The asymptotic behaviour of the potentials is crucial for the
zero-mode.  To see this, we assume ${\rm e}^{i\omega t}$
time-dependence for $\psi_k$ and $\phi_k$, which converts
Eqs.~(\ref{master wave pde}) and (\ref{RS master wave pde}) into
Schr\"odinger-type eigenvalue problems with energy parameter $E =
\omega^2$. From elementary wave mechanics, it is plausible for the
RS potential to support a positive-energy normalizable bound
state, because there are both an attractive delta function and an
asymptotically positive potential. Indeed, such a bound state does
exist with $\omega = \pm k$; this is the zero-mode responsible for
reproducing General Relativity on the brane. By the same token,
the fact that the asymptotic ES potential vanishes strongly means
that it \emph{cannot} support a normalizable bound state with
$\omega^2 > 0$. There is {\em no Randall-Sundrum type zero-mode in
the Einstein-static brane-world.} The vanishing of the potential
reflects the fact that the horizon is perfectly transparent to
gravity waves, so it is fair to say that the delocalization of
brane gravity is entirely due to the bulk black
hole.\footnote{This conclusion is reinforced by the consideration
of an ES brane embedded in pure anti-deSitter space,
\emph{without} a black hole. In that case, the 5D scalar
fluctuation spectrum includes a stable brane-localized mode
\cite{tanaka}, in line with our claim that the horizon is
responsible for the delocalization of brane gravity.}

\begin{figure}
\begin{center}
    \includegraphics[width=\columnwidth]{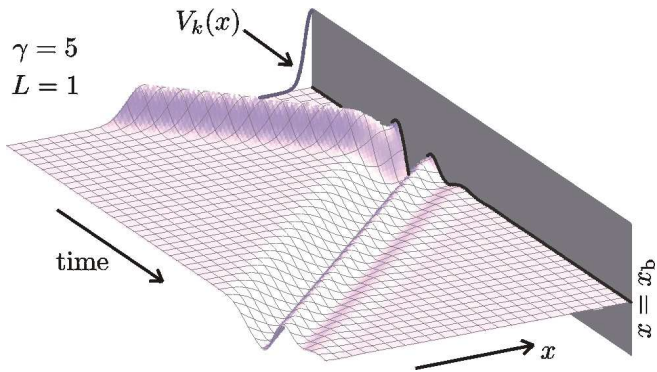}
\end{center}
\caption{Numerical scattering experiment: a Gaussian gravity wave
pulse collides with an Einstein-static brane.}\label{fig:profile}
\end{figure}

Is there a tachyonic instability with $\omega^2<0$ in the ES
brane-world? This cannot be determined from inspection of the
potential in Fig.~\ref{fig:comparison}. One way to answer this is
to conduct `scattering experiments' where the wave
equation~(\ref{master wave pde}) is solved
numerically~\cite{Seahra:2005wk}. As initial data, we choose a
Gaussian pulse moving towards the brane at $t = 0$.
Figure~\ref{fig:profile} shows a rather clean scattering event
where the pulse strikes the brane, is reflected, and then
propagates to infinity. At late times, the brane geometry reverts
to its background configuration as the fluctuation dies away. This
suggests that there is no linear instability in the system, and
that none of the energy in the original pulse becomes trapped on
the brane. Experimenting with a wide variety of incident signals
suggests that both of these conclusions are generic and
independent of initial data.

More inferences can be drawn from the late-time gravity wave
signals on the brane in these scattering experiments, which are
shown in Fig.~\ref{fig:brane signal}. We consider Gaussian
incident pulses with identical characteristics in both the RS and
ES (with $k=L=1$) scenarios, and directly compare the resulting
waveforms. The RS signal is very well approximated by $\phi_k
\approx \text{Re}(A{\rm e}^{i\omega t})$, where $A$ is a complex
constant.  A least-squares fitting gives $\omega = 1.00$, which
implies that we are seeing stable zero-mode oscillations with
$\omega = k$.  On the other hand, the ES signal is exponentially
damped, in agreement with the behaviour seen in
Fig.~\ref{fig:profile}.  The waveform is again well-described by
$\psi_k \approx \text{Re}(A{\rm e}^{i\omega t})$, but with a
complex frequency $\omega = 0.848 + 0.256\,i$.  Hence as $t
\rightarrow \infty$, some of the energy in the incident pulse
remains localized on the RS brane while the ES brane radiates it
all away. This is a direct numerical confirmation of the claim
made above: the brane in the RS scenario supports a stable
zero-mode, while the brane in the ES scenario does not.
\begin{figure}
\begin{center}
    \includegraphics{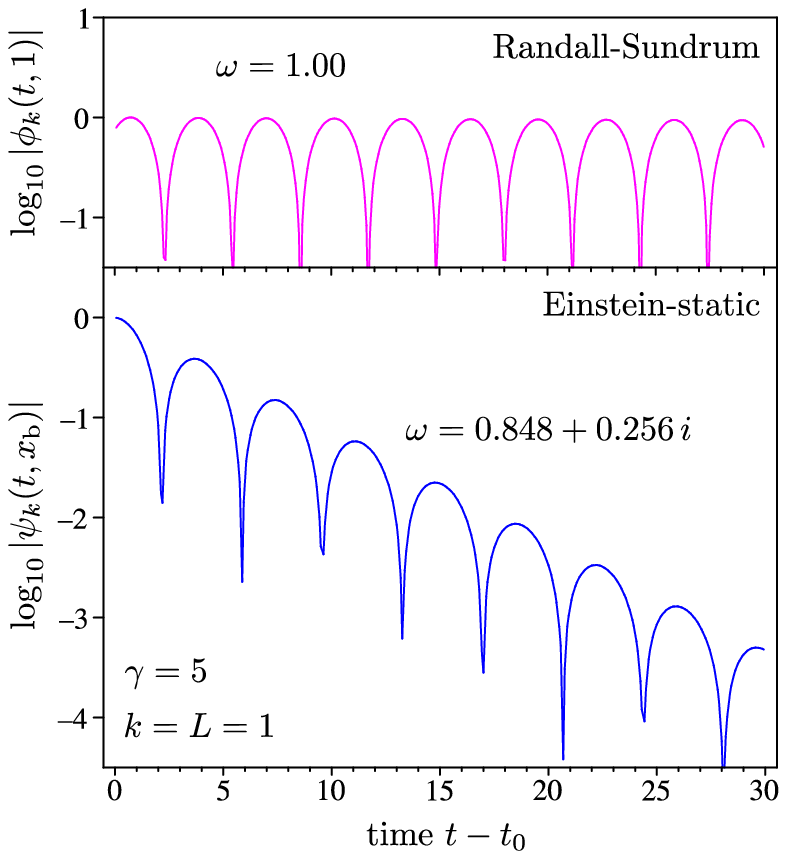}
\end{center}
\caption{Brane gravity wave signals induced by the scattering of
identical Gaussian pulses in the Randall-Sundrum and
Einstein-static ($\gamma = 5$) scenarios, with $k = 1$. The values
of $\omega$ are obtained by fitting a $\text{Re} ( A {\rm
e}^{i\omega t} )$ template to each waveform.}\label{fig:brane
signal}
\end{figure}

\subsection*{Quasinormal modes}

The removal of the zero-mode by the bulk black hole is our main
result, but the complex-frequency oscillations exhibited in the ES
brane-world deserve some investigation. Such behaviour is
reminiscent of the familiar `ringdown' waveform from black hole
perturbation theory, which is a direct consequence of the
existence of so-called `quasinormal modes' (QNMs). These are
solutions of the relevant master wave equation subject to purely
outgoing boundary conditions, and are characteristic of systems
where energy can be lost to infinity.  QNMs are described by a
discrete set of complex frequencies $\omega_n$ with
$\text{Im}\,\omega_n > 0$. Hence, QNMs are exponentially damped in
time. These modes are naturally interpreted as metastable bound
states or scattering resonances of the potential, i.e., `almost
trapped' waves \cite{Taylor}.

It has recently been demonstrated that the brane in the RS
scenario supports QNMs~\cite{Seahra:2005wk}, which are discrete
modes embedded within the Kaluza-Klein continuum of massive modes.
Thus it is perhaps not surprising that the ES brane exhibits
quasinormal ringing.  Indeed, one might have expected QNMs from
the fact that we are really considering a kind of modified 5D
black hole perturbation problem with a boundary.  The novel
feature is that the ES late-time gravity wave signals are {\em
dominated} by QNM oscillations. This is in contrast to the RS
scenario, where the zero-mode usually obscures the quasinormal
ringing. Since these scattering resonances are of crucial
importance to actual gravity wave signals, we calculate the QNM
spectrum for the ES brane-world.

Our method for finding the QNMs, which is discussed in detail
elsewhere~\cite{other_paper}, relies on the series solution of
Eq.~(\ref{master wave pde}) in the frequency domain.  Assuming
such a solution satisfies both the brane boundary condition
Eq.~(\ref{boundary condition}) and the outgoing wave condition,
\begin{equation}
    \psi_k \sim {\rm e}^{i\omega(t+x)} \text{ as } x \rightarrow
    -\infty\,,
\end{equation}
results in an infinite-order polynomial in $\omega$. Truncation of
this polynomial at some order $N$ gives a finite number of complex
solutions $\omega^{(N)}_n$. Frequencies that are stable in the $N
\rightarrow \infty$ limit are the QNM frequencies of the system.

\begin{figure}
\begin{center}
    \includegraphics{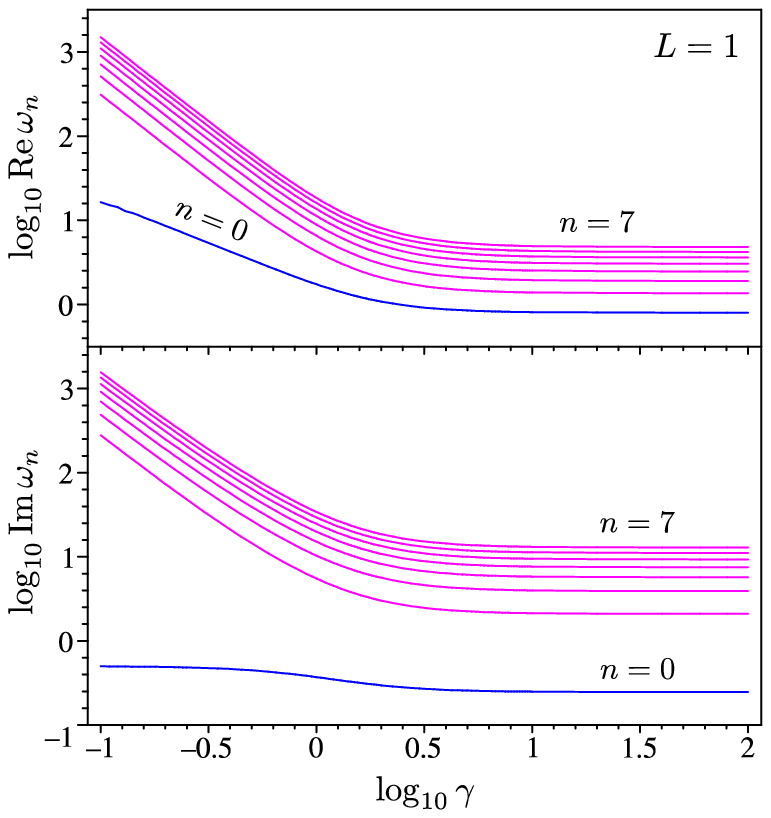}
\end{center}
\caption{Real (\emph{top}) and imaginary (\emph{bottom}) parts of
the first eight quasinormal frequencies in the ES
brane-world.}\label{fig:QNMs}
\end{figure}

We apply this method to calculate the first eight quasinormal
frequencies of the ES brane in the $L=1$ case, as functions of
$\gamma$, and the results are plotted in Fig.~\ref{fig:QNMs}.
Frequencies are labelled in order of increasing modulus, the
smallest being the fundamental mode ($n=0$) and the others being
the overtones ($n=1, \ldots, 7$).  We can check the validity of
the results by examining the value of the fundamental frequency
for $\gamma = 5$:
\begin{equation}
    \omega_0 = 0.8488226912 + 0.2563765163 \,i,
\end{equation}
which is in excellent agreement with the best-fit frequency in the
bottom panel of Fig.~\ref{fig:comparison}.

Figure~\ref{fig:QNMs} shows that the QNM frequencies approach
constant values as $\gamma$ becomes large (i.e.~for small black
holes). This can be understood by examining the leading order
behaviour as $\gamma \rightarrow \infty$ of the master
equation~(\ref{master wave pde}) and boundary
condition~(\ref{boundary condition}). In this limit, $r \le \rb
\approx \sqrt{2}$, and we can neglect terms of order
$r^2/\gamma^2$. This results in the asymptotic wave equation,
\begin{equation}\label{large gamma pde}
-\frac{\di^2 \psi_k}{\di t^2} = -\left( g \frac{\di}{\di r}
\right)^2\! \psi_k + \frac{g [(4k^2 + 11)r^2 + 9 ]}{4 r^4}
\psi_k\,,
\end{equation}
where $g(r) = 1 -1/r^2$.  The boundary condition reduces to
\begin{equation}\label{large gamma bc}
\left[ \frac{\di \psi_k}{\di r} - \frac{3}{2\sqrt{2}} \psi_k
\right]_{r = \sqrt{2}}=0\,.
\end{equation}
Both equations are independent of $\gamma$, and thus the QNM
frequencies should also be independent of $\gamma$ in the $\gamma
\gg 1$ limit, as confirmed in Fig.~\ref{fig:QNMs}. We also find
that the frequencies are evenly spaced for large $\gamma$, and are
well approximated by:
\begin{equation}
    \omega_n \approx 0.78 + 0.28 \, i + (0.57 + 1.81 \, i ) n\,.
\end{equation}
The root mean square error between this relationship and the
calculated frequencies is 0.02 at $\gamma = 100$.

In the small $\gamma$ limit, since $r>1$, we can neglect terms of
order $\gamma^2/r^2$. Writing $\psi_k(t,r) = {\rm e}^{i\omega t}
\Psi_k(r)$, the wave equation for $\gamma\ll1$ is
\begin{equation}\label{small gamma pde}
-\gamma^4\omega^2 \Psi_k = -\left( h \frac{d}{d r} \right)^2\!
\Psi_k + \frac{h(15 r^4+9)}{4r^4} \Psi_k\,,
\end{equation}
where $h(r) = (r^4 - 1)/r^2$. In this equation, $\gamma^2$ and
$\omega$ are degenerate because they only appear in the product
$\gamma^2\omega$. This suggests that the QNM frequencies of the
system should obey a power-law scaling $\omega_n \propto
\gamma^{-2} \approx \kappa $ for $\gamma \ll 1$. This is indeed
true for the case of scalar \cite{Horowitz:1999jd},
electromagnetic, and gravitational \cite{Cardoso:2003cj}
fluctuations of a pure Schwarzschild-AdS black hole with no brane
around it. However, the brane boundary condition for $\gamma \ll
1$,
\begin{equation}
\Psi'_k (\rb) = \frac{3\gamma}{2\sqrt{2}} \Psi_k(\rb)\,, \quad \rb
\approx \frac{\sqrt{2}}{\gamma}\,,
\end{equation}
breaks the degeneracy, and we should not necessarily expect that
$\omega_n \propto\gamma^{-2}$ for $\gamma \rightarrow 0$. But in
fact, we do see such a scaling for the overtone frequencies in
Fig.~\ref{fig:QNMs}:
\begin{equation}\label{asymptotic overtones}
\omega_n \approx \Omega_n \gamma^{-2} \text{ for } \gamma \lesssim
0.14\,, \,\, n = 1,2,\ldots\,,
\end{equation}
where $\Omega_n$ is a complex constant determined numerically.
For example, by performing a fit between $\gamma = 0.10$ and 0.14,
we find $\Omega_1 = 3.109 + 2.814 \, i$.  The goodness of the fit
can be assessed by looking at the RMS discrepancy between the
logarithms of the calculated and approximate frequencies.  For
$n=1$, this error is $10^{-4}$ and $2 \times 10^{-3}$ for the real
and imaginary parts, respectively.

The small-$\gamma$ behaviour of the fundamental mode is quite
different.  We find
\begin{equation}\label{asymptotic fundemental}
\omega_0 \approx 1.68 \gamma^{-1} + 0.497 \,i \text{ for } \gamma
\lesssim 0.14\,,
\end{equation}
i.e., the real part appears to scale like $\gamma^{-1}$ while the
imaginary part approaches a constant.  The discrepancy in the
asymptotic behaviour of the overtones (\ref{asymptotic overtones})
and the fundamental mode (\ref{asymptotic fundemental}) would seem
to suggest that the latter is more sensitive to the boundary
condition.

Attempts to test these relations for much smaller values of
$\gamma$ are constrained by computing speed, since for $\gamma
\rightarrow 0$, we have $\rb \rightarrow \infty$, which is a
singular point of the master wave equation~(\ref{master wave pde})
in the frequency domain. Hence the series solution for $\psi_k$
becomes poorly convergent at the brane, which means we must retain
unreasonably many terms to get accurate QNM frequencies.

We have also calculated QNM frequencies for different values of
$L$, i.e., for perturbations on different spatial scales on the
brane. The results for $\gamma = 5$ are shown in Fig.~\ref{fig:L
varying}. The general trend is that the real parts of the
frequencies increase with $L$ (i.e., for smaller scales), while
the imaginary parts remain roughly constant or decrease slightly.
This is in keeping with results obtained for the scalar QNMs of a
4D Schwarzschild-AdS black hole with no
brane~\cite{Horowitz:1999jd}.
\begin{figure}
\begin{center}
    \includegraphics{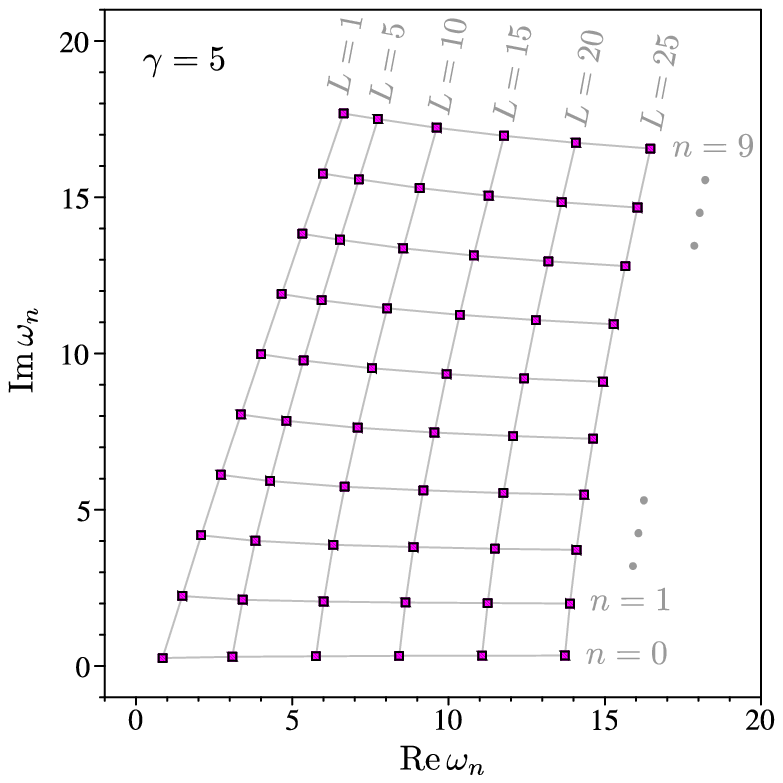}
\end{center}
\caption{The first ten quasinormal frequencies of the $\gamma = 5$
ES brane for various values of $L$.}\label{fig:L varying}
\end{figure}

\subsection*{Kaluza-Klein masses?}

In the RS scenario, the various modes of the 5D graviton can be
labelled by an effective 4D Kaluza-Klein (KK) mass. This can be
understood as follows: The RS potential in (\ref{RS master wave
pde}) can be re-written as
\begin{equation}
U_k(z)= k^2 + \overline{U}(z)\,,
\end{equation}
where $\overline{U}$ is independent of the spatial scale $k$. With
$\phi_k(t,z)={\rm e}^{i\omega t} \varphi_k(z)$, we find the
dispersion relation
\begin{equation}\label{disp}
m^2 = E^2 - p^2, \quad E = \omega/\ell, \quad p = k/\ell\,,
\end{equation}
where $m^2\ell^2$ is the eigenvalue of $-(d/dz)^2 +
\overline{U}(z)$.  To a stationary brane observer, $E$ is the
energy of the mode while $p$ is its 3-momentum. Hence, $m =
\sqrt{E^2 - p^2}$ is naturally interpreted as the effective 4D
graviton mass according to the standard KK paradigm.

By contrast, in the ES brane-world, the spatial scale $k$ can no
longer be separated from the potential $V_k$, and no simple
position-independent dispersion relation like (\ref{disp}) can be
found. At best, we can define a local dispersion relation on the
brane:
\begin{equation}\label{mass def'n}
\tilde{m}^2 = \tilde{E}^2 - \tilde{p}^2, \quad \tilde{E} =
\frac{\omega}{\Rh \sqrt{\fb}}, \quad  \tilde{p} = \frac{k}{\Rh\rb}
\,.
\end{equation}
Note that in the shortwave approximation $k \gg 1$, $\tilde{E}$
and $\tilde{p}$ are the mode's energy and 3-momentum as measured
by comoving brane observers, respectively, and $\tilde{m}$ is the
effective mass of the mode.

\begin{figure}
\begin{center}
    \includegraphics{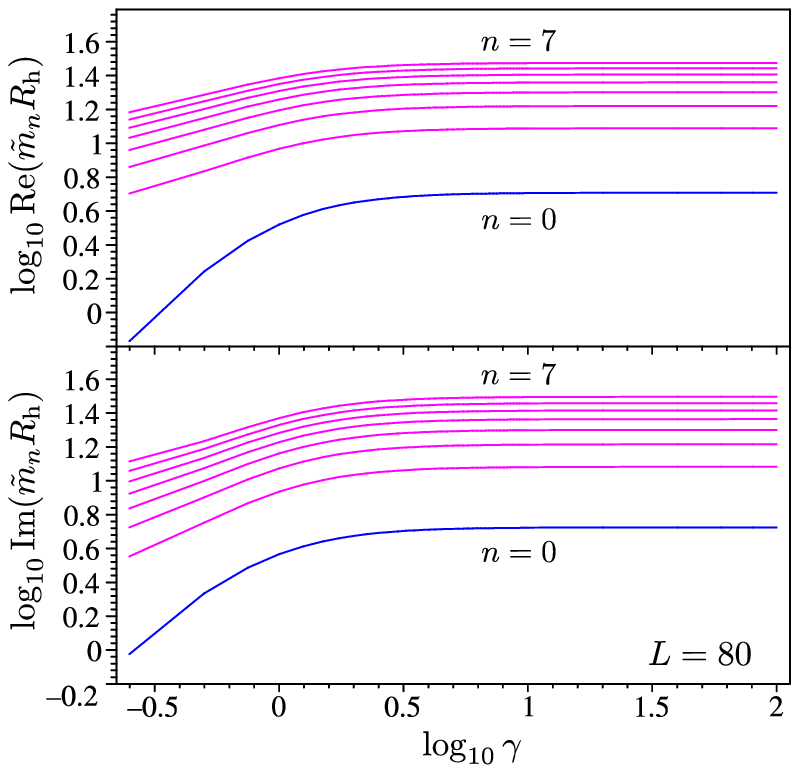}
\end{center}
\caption{Effective 4D masses, as measured by brane observers, of
the first eight QNMs in the $L = 80$ case}\label{fig:masses}
\end{figure}

What are the masses of the QNMs we calculated above?  We continue
(\ref{mass def'n}) into the complex plane in order to define a
complex mass for each QNM. Then, $\text{Re}\, \tilde m_n$ is
interpreted as the conventional mass of the resonance, while
$(\text{Im}\,\tilde m_n)^{-1}$ is roughly its half-life. Our
results for $\tilde{m}_n$ for $L=80$ are shown in
Fig.~\ref{fig:masses}. Interestingly, the upper panel suggests a
mass gap between the fundamental mode and the overtones for small
$\gamma$, i.e., for large black hole effects. This is reminiscent
of a de Sitter brane-world~\cite{Garriga:1999bq} (although in the
de Sitter case, the gap is between the zero-mode and the KK
continuum).

\subsection*{Generalizations}

One can generalize the ES brane-world model by making the bulk
more complicated, letting the brane move, or both.  Will our
results carry over to these new situations?

First consider the addition of matter fields in the bulk. These
may take the form of dilatons, moduli fields, supergravity form
fields, etc. Stationary, spherically symmetric solutions sourced
by such matter and featuring an event horizon are classified as
``dirty black holes''. Scalar wave propagation in 4D dirty black
hole spacetimes shows~\cite{Medved:2003rg} that the potential in
the master wave equation vanishes at the horizon, just as in
Eq.~(\ref{master wave pde}). Now, the key reason that the
zero-mode becomes delocalized in the ES model is the fact that
$V(\Rh) = 0$.  Hence, if we make the natural assumption that the
behaviour seen in Ref.~\cite{Medved:2003rg} generalizes to spin-2
fields and higher-dimensional backgrounds, we conclude that the
zero-mode will remain delocalized if the 5D Schwarzschild-AdS bulk
is replaced with a dirty black hole. Indeed, we can go even
further by conjecturing that \emph{any static brane located
outside a stationary Killing horizon cannot support a normalizable
zero-mode}, precisely because any such horizon be will completely
transparent to bulk gravity waves.

Next, consider brane motion in a Schwarzschild-AdS bulk, $\rb =
\rb(t)$ as in model of brane cosmology.  Do the QNM frequencies we
have calculated tell us anything about the behaviour of
cosmological perturbations? Fluctuations with $|\omega| \gg
H\equiv \dot{r}_\text{b}/\rb$ do not `feel' the expansion of the
universe and effectively `see' the brane as stationary, i.e., the
characteristic timescale of the perturbation is much larger than
the characteristic timescale of the cosmological dynamics. Hence,
we can expect the QNMs calculated for a static brane to be
approximate solutions for the gravity waves around a moving brane
if $1/|\omega_n|$ is much shorter than the Hubble time. Such QNMs
will dominate that part of the bulk gravity wave spectrum. Hence,
\emph{the high frequency $\omega \gg H$ component of the bulk
gravity wave spectrum on a cosmological brane should be dominated
by the metastable bound state resonances of the corresponding
static brane}.

Such an approximation has the potential to greatly simplify the
thorny problem of brane cosmological perturbations, but there is a
significant caveat. The QNMs we have calculated are only for one
brane position, i.e., on the photon sphere. In order to have a
complete picture, we need to know the QNM frequencies for static
branes over a range in $r$, corresponding to the addition of
matter on the ES brane. The calculation of these frequencies is
the subject of a separate paper~\cite{other_paper}.

\subsection*{Conclusions}

We have considered static pure-tension branes surrounding a 5D
bulk black hole.  By studying the tensor perturbations, we have
seen that the bulk Killing horizon causes the brane's zero-mode to
become delocalized. In other words, the gravitational field of the
black hole makes it impossible for the brane to support a
normalizable bound state. However, we also found that the brane
supports a discrete spectrum of metastable bound states, or
quasinormal modes, as in the Randall-Sundrum scenario. Using a
series solution of the master wave equation, we have calculated
the quasinormal frequency spectrum. We discussed why the massive
mode Kaluza-Klein decomposition common in other brane-world models
does not work in the current problem, but then showed how one
could define an effective local mass measured by brane observers.
The locally defined mass shows a gap between the fundamental and
overtone modes.

Our results are expected to generalize in several important ways.
We expect that whenever there is a stationary Killing horizon in
the bulk, a surrounding brane cannot support a normalizable bound
state. Furthermore, we expect that the high frequency bulk gravity
wave spectrum on a moving brane will be well represented by a sum
over the quasinormal resonances of the corresponding static brane.

\subsection*{Acknowledgements}

SSS is supported by NSERC, CC and RM by PPARC. We thank K Koyama,
A Mennim, and D Wands for discussions.



\end{document}